# DISTRIBUTED AND BIG DATA STORAGE MANAGEMENT IN GRID COMPUTING

Ajay Kumar[1] and Seema Bawa[2]

[1]Department of Computer Science and IT, Mewar University, Chittorgarh, INDIA
`ajaycpp@gmail.com`
[2]Department of Computer Engineering, Thapar University, Patiala, INDIA
`seema@thapar.edu`

*Abstract*

***Big data storage management is one of the most challenging issues for Grid computing environments, since large amount of data intensive applications frequently involve a high degree of data access locality. Grid applications typically deal with large amounts of data. In traditional approaches high-performance computing consists dedicated servers that are used to data storage and data replication. In this paper we present a new mechanism for distributed and big data storage and resource discovery services. Here we proposed an architecture named Dynamic and Scalable Storage Management (DSSM) architecture in grid environments. This allows in grid computing not only sharing the computational cycles, but also share the storage space. The storage can be transparently accessed from any grid machine, allowing easy data sharing among grid users and applications. The concept of virtual ids that, allows the creation of virtual spaces has been introduced and used. The DSSM divides all Grid Oriented Storage devices (nodes) into multiple geographically distributed domains and to facilitate the locality and simplify the intra-domain storage management. Grid service based storage resources are adopted to stack simple modular service piece by piece as demand grows. To this end, we propose four axes that define: DSSM architecture and algorithms description, Storage resources and resource discovery into Grid service, Evaluate purpose prototype system, dynamically, scalability, and bandwidth, and Discuss results. Algorithms at bottom and upper level for standardization dynamic and scalable storage management, along with higher bandwidths have been designed.***

## 1. INTRODUCTION

A Grid is a collection of machines, sometimes referred to as "nodes", "resources", "members", "clients", "hosts", "engines", and many other terms[1]. Grid environments are increasingly being used by applications such as particle physics, climate modelling, whether forecasting or astrophysics. They all contribute any combination of resources to the Grid as a whole. These applications make use of unique, high-end supercomputers and large-scale data storage-systems and produce large multi-dimensional data sets. This type of work is increasingly being performed in collaborative efforts between geographically distributed scientists and organizations, utilizing shared resources that are also widely distributed. Network based storage systems such as Network Attached Storage (NAS) [9] and Storage Area Network (SAN) [3] offer a robust and easy method to control and access large amounts of storage. However, with the steady growth of client access demands and the required data sizes, it is a challenge to design an autonomous, dynamic, large-scale and scalable storage system which can consolidate distributed storage resources to satisfy both the bandwidth and storage capacity requirements [2].

Some resources may be used by all users of the Grid while others may have specific restrictions. In this research paper we are focusing to achieve dynamicity and scalability for large scale data storage system in Grid environments. The Dynamic and Scalable Storage Management (DSSM) architecture to organize Grid Oriented Service (GOS) devices into a large –scale and geographically distributed data storage system to meet the requirements imposed by all kinds of Gris applications [2].

## 2. Data Management in Grid Computing

In this section we explain the evolution of the data management within the Grid and investigate the various implementations of the data Grids that are in place today as well as those currently emerging.

### 2.1. General Traditional Data Management

Electronic data management has a long and rich history dating back to the 1950s many data management systems have tried to make their way into the mainstream of information technology, some more successfully than others; hierarchical, network, object, and in-memory are only a few examples. The most successful data management system has been the relational data management technology [10]. We will start at this point to look at its development and what has made it succeed and also how far forward it has moved in comparison to any other data management system.

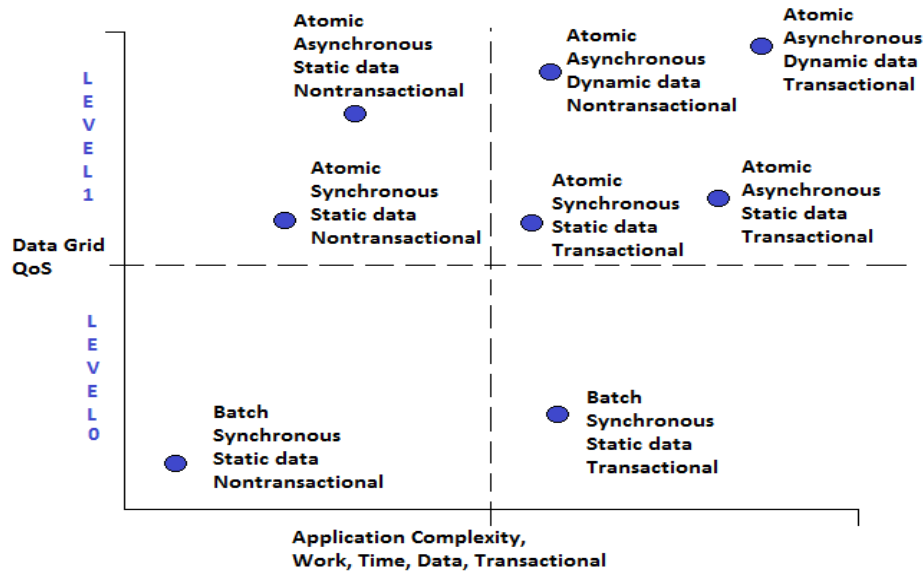

Fig. 1: Date Grid Quality of Service vs. Application Demand and Requirement [1]

## 3. Related Work

### 3.1. Proposed Architecture

The proposed architecture named "Dynamic and Scalable Storage Management (DSSM)" to organize Grid storage devices into a large-scale and geographically distributed storage system to meet the requirements imposed by all kinds of Grid applications. The DSSM divides Grid storage devices into multiple geographically distributed domains to facilitate the data access locality [2][5]. The architecture consists of two levels. The bottom level adopts multicast to achieve dynamic, scalable, and self-organized physical domains. The method significantly simplifies the intra-domain storage resource management. The upper level is a virtual domain that consists of geographically distributed and dynamic agents selected from each physical domain.

## 3.2. Data Locality

Data locality is a measure of how well data can be selected, retrieved, compactly stored, and reused for subsequent accesses [2]. In general, there are two basic types of data locality: temporal and spatial. Temporal locality denotes that the data accessed at one point in time will be accessed in the near future. Temporal locality relies on the access pattern of different applications and can therefore change dynamically [6][7]. Spatial locality defines that the probability of accessing data is higher if the data near it was just accessed (e.g. pre-fetch). Unlike the temporal locality, spatial locality is inherent in the data managed by a storage system, and is relatively more stable and does not depend on applications, but rather on data organizations which is closely related to the system architecture Reshaping access patterns can be employed to improve temporal locality [2]. Data reorganization is normally adopted to improve spatial locality. Many research efforts have been invested in exploiting the impact of access pattern and data organization of applications on the data locality to achieve performance gains. Fig.-3.1 illustrates the DSSM architecture.

## 3.3. Domain Division

Choosing a suitable criterion to form GOS nodes into domains is an important factor of the DSSM architecture. DSSM employs the distance criterion in a geographical way to divide GOS nodes into multiple domains to facilitate the data access locality and limit the amount of communication traffic seen by each node with performance guarantee. GOS devices belonging to the same geographical area (e.g. the same enterprise or the same LAN) are formed into a domain, because the GOS devices in the area are normally placed geographically close [11][12]. GOS nodes divide into multiple domains based on:

- Distance
- Quantitative (bandwidth)
- Qualitative (performance analysis)

With a domain based GOS Network, the scalability is achieved from a two-tiered architecture, namely, intra-domain and inter-domain. Within a domain, each GOS device is equal in functions and capabilities, and can leave or join the domain dynamically. Any node can communicate with anyone else directly. Each domain selects a domain agent from the domain members to cooperate with other domains according to domain formation algorithm [2][13].

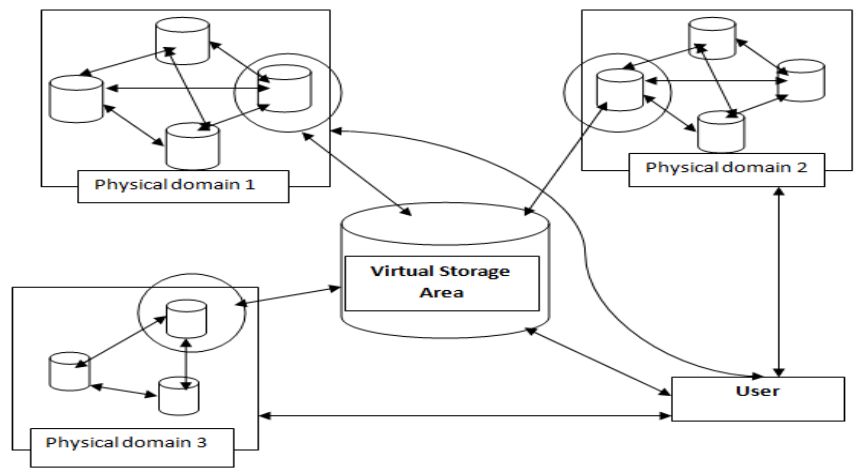

Fig. 2: The DSSM Architecture

## 4. DATA MANAGEMENT IN GRID COMPUTING

The DSSM architecture consists of two levels. The bottom level adopts multicast to achieve dynamic, scalable, and self-organized physical domains. The upper level is a virtual domain that consists of geographically distributed and dynamic GOS agents selected from all domains. The algorithm consists of knowledge of the neighbours of each node in a domain to organize the domain members and select a domain agent from all candidate members. Several distributed domain formation algorithms have been devised over the years. One is the lowest-ID algorithm. The other is the highest-connectivity (degree) algorithm.

### 4.1. Algorithm Formulation at Bottom Level

DSSM architecture organized as a scalable storage system. Assign with a multicast IP address to provide a single entry point the storage space. Each GOS device maintains an Adjacent Information Table (AIT) which keeps a list of all active GOS devices with their related resource information. Information consists of IP address, reminder storage capacity, processing power. A data structure is adopted to describe the entry of AIT. The data structure should be maintained in memory so that they can be accessed with little overhead. The data structure takes 32 bytes per entry.

For a physical domain which has 1000 GOS devices, it takes only

1000*32 = 32000 bytes

to track the whole physical domain.

Compared with the size of main memory, the storage capacity of AIT is negligible.

Table 1: Adjacent Information Table (AIT)

| Field | Size | Key Value |
| --- | --- | --- |
| AIT ID | 4 bytes | Alphanumeric |
| IP address | 32-bits | Alphanumeric |
| Storage capacity | 8 bytes (in MB) | Floating points |
| Processing power | 8 bytes (in MHz) | Floating points |

**Algorithm#1: When a new GOS device wants to join the domain:**

**Step1: Send "JOIN"** (the device sends a one hop probing multicast "JOIN" request of its coming and the corresponding local resource information such as the processing power, storage capacity, and waits for the response of other online GOS devices).
**Step2: Receive "JOIN"** request (the online GOS devices which receive the "JOIN" request add the oncoming device's information to their AIT).
**Step3: Send ack "ACCEPT"** (online GOS devices send an "ACCEPT" ack back to the probing GOS device using unicast)
**Step4: Create AIT** (the oncoming device constructs its own AIT in terms of the ack messages.)

**Algorithm#2: Leaving of a GOS device**

DSSM is based on stable and trusted GOS devices; the leasing of a GOS device is normally caused by maintenance, upgrade and other reasons.

**Step1: Send "LEAVE"** (multicast leave message in the domain)

**Step2: Receive "LEAVE"** (once the online GOS devices receive the message, the devices delete the leaving device's information)

**Step3: Update AIT**

**Step4: Repeat Step2** (if message is not received then the device is assumed to have failed and other reminder GOS devices delete the device from their AIT).

## 4.2. Algorithm Formulation at Upper Level

**Algorithm#3: Select agent from particular physical domain**

**Step1: Select MAX[PP]** (select highest processing power agent within a domain)
**Step2: Repeat Step1** (if fail to select highest processing power agent)
**Step3: Compare agent to another GOS**
**Step4: Repeat Step1** (highest processing power will always reselect as an agent)
**Step5: In case of same processing power unchanged.**

## 5. STORAGE RESOURCE MONITORING AND DISCOVERING (SRMD)

Grid users are interacting with SRMD [10][15]. Within the grid environment, the storage users can achieve the following functions [9][14][16]:-

1)    Obtain a certificate from the certificate authority.

2)    With the help of SRMD, search and discovery appropriate storage service

3)    Access the corresponding storage resource transparently.

### 5.1 Structure of SRMD

```
1:http:/192.168.16.10:8080/srmd/services/storservice
2:http:/192.168.16.12:8080/srmd/services/mgtservice
3:http:/192.168.16.20:8080/srmd/services/secservice
4:http:/192.168.16.10:8080/srmd/services/comservice
   -    - - - - - -- - - -- - - - - - - -- - - - - -- -- - -
   -       - -- - - - -- - - - - - - - -- - - -- - - -- -  -
   -      - -- --- - - - - - - - - - - - - - - - -- -- - -
```

**Fig 3.1: SRMD structure**

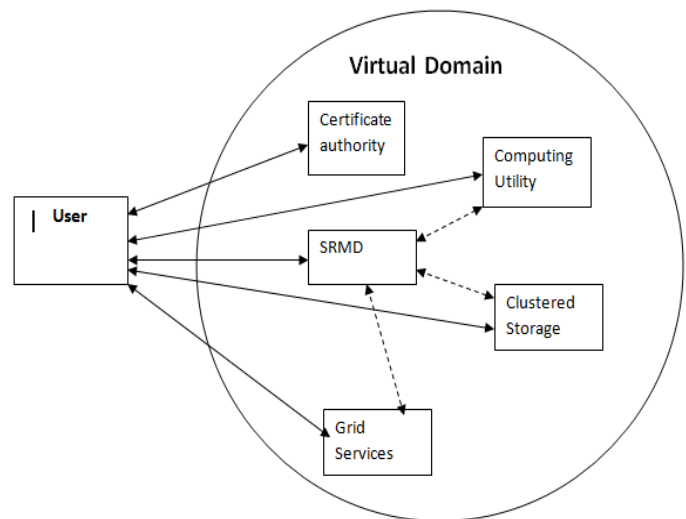

**Fig 3.2: Storage Resource Monitoring and Discovery**

# 6. IMPLEMENTATION AND EXPERIMENTAL EVALUATION

We constructed a prototype with three different Thapar University's networks, one High End Computer Lab (HECL) Network emulator, Software Engineering Lab (SEL) Network and several client machines of PG Hostel, J Hostel's networks. All components were connected through a 1000/100M adaptive switch. Table-1 shows the system configurations of the prototype. The WAN emulator forwards packet at line rate and has user-settable delay and drop probability.

## 6.1. Dynamicity and reliability evaluation

The dynamicity and reliability of the DSSM architecture were measured at two stages. At the first stage, we set the IP address of three GOS devices within one network segment to denote a single physical domain. To illustrate the dynamic scalability and reliability of the architecture, we first configured one GOS device in HECL Network, and then the other two GOS devices joined the domain SEL Network, PG Hostel Network respectively, finally, the three GOS devices continuously left and joined the domain.

Table 2: System Configurations of the Prototypes

| H/W | GOS Device | WAN Emulator and Clients |
|---|---|---|
| CPU | Intel Xeon 2.8GHZ | Intel Pentium IV 2.66 GHZ |
| Memory | 3GB | 512MB |
| NIC | Two Broadcom 100/1000M | Intel(R) PRO/100M |
| Disks | Six IBM FRU 32P0730 | Seagate ST340014A |
| OS | Red Hat(Kernel 2.4.21) | Red Hat (Kernel 2.4.21)/MS Windows 7 |

We repeated the above process for more than 50 times, it did not cause any problems in our experiment. Because all GOS devices in the domain have the same processing power, the first GOS device of the domain is selected as an agent by default even it has no other GOS agents to cooperate with. At the second stage, to simulate two physical domains (HECL Network, SEL Network), we configured two network segments by setting the IP address of the three GOS devices. The first domain consisted of two GOS devices, and the second domain had only one GOS device.

## 6.2 Bandwidth Evaluation

This paper focuses on a dynamic and scalable storage management architecture which may involve hundreds or even thousands of distributed GOS devices. The goal of this work is to support long-distance and bulk data access of large-scale and complex Grid applications. Grid service combines the Web service and WSRF to provide a service based Grid environment that enables heterogeneous environments to be integrated and reconciled to accomplish complex tasks [8]. A set of files appropriate range were transferred over the emulated WAN to measure the bandwidth.

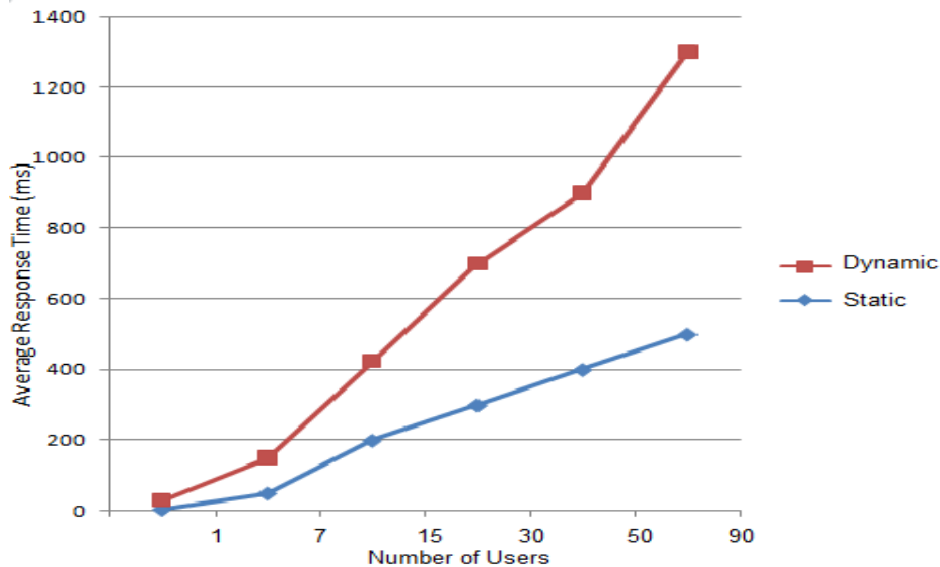

Fig. 3: Average Response Time of the Static and Dynamic Storage System.

## 7. FUTURE WORKS

The first, if hundreds of thousands query requests go to a particular domain agent simultaneously, the agent could become a potential system bottleneck. The reason is that all query traffic going to a domain has to pass through the domain agent. The second, an intelligent domain agent is able to pool the storage resources and keep the load balance among the GOS devices in the domain. It takes some additional computing overhead to do this job. The third one is the worst scenario. When a GOS agent is serving data access to users, a large number of query requests going to the agent will heavily overload the agent. This scenario should be avoided or alleviated. However, the I/O traffic taken over by each agent will be decreased with the increase in number of domain agents. The probability that most of the I/O traffic goes to a particular domain can be decreased, if the hot data is arranged properly. A dynamic and transparent data replication mechanism which automatically places the data replicas across domains where they are needed is able to alleviate the performance impact on the GOS agents, it is also crucial to the overall performance. The first, it improves the aggregate bandwidth by providing simultaneous data access to multiple data copies. The second, it reduces the latency between the data provider and data consumer by copying the data near the data consumer,

## 8. CONCLUSION

In this research paper, based on an existing Grid environment, we propose and design a DSSM architecture which organizes GOS devices into different domains to enhance the data access locality in terms of geographical and distributed areas. Storage expansion is achieved by simply adding services. The DSSM architecture avoids the hierarchical or centralized approaches of traditional Grid architecture, eliminates the flat flooding of unstructured P2P system, and provides a dynamic storage pool in Grid environment. The proof-of-concept prototype gives useful insights into the architecture behaviour of DSSM architecture.

## ACKNOWLEDGEMENTS

I sincerely acknowledge the guidance rendered to me by Dr. Seema Bawa, Professor, at Computer Science and Engineering Department, Thapar University, Patiala. Words are not enough to describe her invaluable support; inspiration, guidance, encouragement and making me understand easily anything, anytime during the research work.

I would like to thank to Dr. Maninder Singh, Professor, Thapar University, Patiala who really put me in a real technical frame and bent of mind – that allowed me to conceptualize and materialize this concept.

I am deeply indebted to our Sr. Grid Members (Ms. Shashi, Ms. Rajani, Ms. Seemu, Ms. Pankajdeep Kaur, Ms. Ratinder Kaur, Ms. Nidhi) who constantly encouraged me tender my utmost gratitude and appreciation for her invaluable guidance and suggestions.

## REFERENCES

[1]  Michael Di Stefano, Distributed Data Management for Grid Computing, ISBN 0-471-68719-7, 2005.

[2] Yuhui Deng , Frank Wang, Na Helian, Sining Wu, Chenhan Liao (2008) "Dynamic and scalable storage management architecture for Grid Oriented Storage device" Parallel Computing 34 (2008) 17-31.

[3] Arie Shoshani, Alex Sim, Junmin "Storage Resource Managers: Middleware Components for Grid Storage" Gu Lawrence Berkeley National Laboratory Berkeley, California 94720.

[4] B. Yang, H. Garcia-Molina, Improving search in peer-to-peer networks, in: Proceedings of the 22nd International Conference on Distributed Computing Systems (ICDCS'02), Vienna, Austria, July 2002, pp. 5–14.

[5] Yuhui Deng and Frank Wang, "A heterogeneous storage Grid enabled by Grid service". ACM SIGOPS operating systems review, Special Issue: File and Storage Systems 41 (1) (2007).

[6] "The Open Grid Services Architecture", <http://www.globus.org/ogsa/>.

[7] Frederik W. Jansen, Erik Reinhard, "Data locality in parallel rendering", in: Proceedings of the 2nd Eurographics Workshop on Parallel Graphics and Visualisation, 1998, pp. 1–15.

[8] Web Services Resource Framework, <http://www.globus.org/wsrf/>.

[9] Yuhui Deng, "Deconstructing Network Attached Storage systems", aEMC Research China, Beijing 100084, PR China -2008.

[10]  Jason McHugh,  Dallan Quass,  Jennifer Widom "Lore  A Database Managemen t System for Semistructur ed Data" see - http://citeseerx.ist.psu.edu/viewdoc/download?doi=10.1.1.81...pdf.

[11] Ajay Kumar, Seema Bawa, Vishnu Sharma "Dynamic and Scalable Data Storage Management in Grid environments", National Conference on Emerging trend in Engineering and Sciences at Samrat Ashok Technological Institute, (M.P.), India, Dec 2010.

[12] Ajay Kumar, Seema Bawa, " Performance Modeling and Run Time Estimation for Large-Scale Data and Computational Grid", International Conference on Advance in Modeling, Optimization and Computing-2011, IIT Roorkee (U.K.), India, ISBN  81-86224-71-2.

[13]  Monitoring and Discovery System, <http://www.globus.org/toolkit/mds/>.

[14]  F. Wang, N. Helian, S. Wu, Y. Deng, K. Zhou, et al., Grid-oriented storage, IEEE Distributed Systems Online 6 (9) (2005) 1–4.

[15]   D. Talia, P. Trunfio, Towards a synergy between P2P and grids, IEEE Internet Computing 7 (4) (2003) 94–96.

[16]   A. Rowstron, P. Druschel, Storage management and caching in PAST, a large-scale, persistent peer-to-peer storage utility, in: Proceedings of the 18th ACM Symposium on Operating Systems Principles (SOSP'01), 2001, pp. 188–201.

**About Author**

**Ajay Kumar** has completed BCA/MCA and M. E. (SE) programs from IGNOU New Delhi and Thapar University Patiala respectively. Currently he is working as an assistant professor at CSE department, Mewar University, Chittorgarh and pursuing Ph. D. program from Thapar University also. His areas of interests include Grid Computing, Software Engineering, Web services and Network Security. His current research focus on large-scale data management in Cloud Computing environments. In addition to his contribution to C# .Net Developer on various domains also published many research papers.

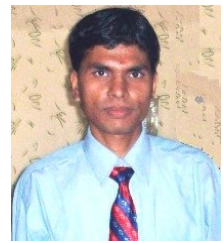

**Dr. Seema Bawa**, an alumni of IIT Kanpur, IIT Kharagpur and Thapar University has demonstrated her intellectual, interpersonal and managerial skills in various domains. Having vast industrial experience of working in IT industry with the role of Project Leader and Project Manager, currently she is a Professor of Computer Science and Engineering and Head of Centre for Information Super Highway at Thapar University, Patiala. Her areas of interests include Parallel and Distributed Computing, Grid Computing, VLSI Testing and Network Management. Her current research focus is on Cloud Computing and Cultural Computing. Along with being a committed teacher and a passionate researcher, she has been actively contributing her services for underprivileged sections of the society.

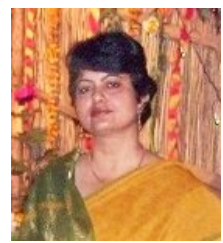